\documentclass{iopjournal}
\usepackage{graphicx}
\usepackage{multirow}
\usepackage{booktabs}
\usepackage{amsfonts}
\usepackage{dcolumn}%
\usepackage{color,ulem}
\usepackage{amsmath}
\usepackage{bm}
\usepackage{calligra}
\usepackage{calrsfs}
\usepackage{mathrsfs}
\usepackage{color}
\usepackage[maxfloats=256]{morefloats}
\usepackage{hyperref} 
\usepackage[pagewise]{lineno}

\begin{document}

\articletype{Paper} 

\title{Spin Josephson diode effect induced by higher-harmonic spin Josephson currents in a diffusive Josephson junction}

\author{Shin-ichi Hikino$^{*}$}

\affil{$^*$National Institute of Technology (KOSEN), Fukui College, Sabae, Fukui 916-8507, Japan}

\email{hikino@fukui-nct.ac.jp}


\begin{abstract}
We theoretically investigate the spin Josephson diode effect (SJDE) in a diffusive Josephson junction with a Rashba metal layer under a ferromagnetic exchange field. 
Within the quasiclassical Green's function framework, we derive analytical expressions for the first- and second-harmonic spin Josephson currents. 
The interplay between Rashba spin–orbit interaction and the exchange field breaks inversion and time-reversal symmetries, generating additional cosine terms in the spin current–phase relations and a finite $\varphi_{0}$ phase shift. 
This phase shift induces an intrinsic asymmetry between forward and backward spin currents, leading to the SJDE without an external magnetic field. 
Numerical results show that the efficiency decreases with increasing metal thickness due to suppression of the second-harmonic component, while its dependence on spin–orbit interaction strength reflects competing effects between phase shift enhancement and harmonic suppression. 
These findings demonstrate that the interplay between harmonic components provides a mechanism for nonreciprocal spin transport without requiring suppression of spin-singlet correlations.
\end{abstract}

\section{Introduction}
Nonreciprocal transport has attracted considerable attention as a fundamental phenomenon in modern electronics. 
In general, nonreciprocity arises from the simultaneous breaking of spatial inversion and time-reversal symmetries, 
leading to asymmetric transport properties with respect to the current direction. 
A representative example is the semiconductor diode based on a p–n junction, which enables directional control of charge flow. 
Such nonreciprocal devices constitute essential building blocks in contemporary electronic technologies. 
Extending this concept to superconducting systems has recently emerged as an important research direction.

Recently, Josephson junctions composed of two $s$-wave superconductors separated by non-superconducting materials~\cite{josephson, degennes, likhalev, golubov-rmp, buzdin-rmp, bergeret-rmp} have been shown to exhibit nonreciprocal transport in the Josephson current, known as the Josephson diode effect (JDE)~\cite{chen-prb98, baumgartner-nt17, pal-nt17, souto-prl129, wu-nature604, zhang-prx12, ciaccia-prr5, tian-apl123, gupta-nc14, greco-apl123, chiles-nlt23, wei-prb108, hu-prl130, kim-nc15, li-acs18, zhang-pra21, coraiola-acs18, huang-apl125, fracassi-apl124, hikino-jpsj94}. 
This effect is characterized by an asymmetric Josephson current, where the critical currents in the forward and backward directions become unequal ($I_{\rm c+} \neq |I_{\rm c-}|$), reflecting an asymmetric current--phase relation $I(\theta) \neq -I(-\theta)$, where $\theta$ is the superconducting phase difference between the two superconductors. 
To realize the JDE, it is essential to simultaneously break both inversion symmetry ($\mathcal{P}$) and time-reversal symmetry ($\mathcal{T}$), 
since $\mathcal{P}$ prevents directional asymmetry, while $\mathcal{T}$ enforces the antisymmetric relation $I(\theta) = -I(-\theta)$. 

When both symmetries are broken, the current--phase relation (CPR) acquires an additional phase shift, leading to a so-called $\varphi_{0}$-junction~\cite{buzdin-prl1, goldobin-prl107, bergeret-epl, szombati-natp, mayer-natc, strambini, hikino-ph}. 
It is important to distinguish the $\pi$ phase shift in superconductor/ferromagnet/superconductor junctions from the $\varphi_{0}$ phase shift~\cite{golubov-rmp, buzdin-rmp, bergeret-rmp, ryazanov-prl, kontos-prl}. 
While the former stems from the spatial oscillation of the superconducting order parameter induced by the exchange field in the ferromagnet, the latter represents a fundamentally different symmetry-breaking mechanism. 
The $\varphi_{0}$ phase shift originates from the simultaneous breaking of $\mathcal{P}$ and $\mathcal{T}$, reflecting an intrinsic phase bias in the CPR.
However, the $\varphi_{0}$ phase shift alone is not sufficient to generate the JDE. 
The reason is that the CPR remains a shifted sinusoidal function that preserves the magnitude symmetry between forward and backward currents. 
The CPR in a $\varphi_{0}$ junction is given by $I(\theta)=I_{\rm c}\sin(\theta+\varphi_{0})$. 
To realize the JDE, it is essential that the CPR contains higher harmonic components, which introduce both even and odd contributions in the phase difference, resulting in an asymmetric relation $I(\theta) \neq -I(-\theta)$~\cite{chen-prb98, baumgartner-nt17, pal-nt17, souto-prl129, wu-nature604, zhang-prx12, ciaccia-prr5, tian-apl123, gupta-nc14, greco-apl123, chiles-nlt23, wei-prb108, hu-prl130, kim-nc15, li-acs18, zhang-pra21, coraiola-acs18, huang-apl125, fracassi-apl124, hikino-jpsj94}. 
Such nontrivial CPRs generally arise from the simultaneous breaking of $\mathcal{P}$ and $\mathcal{T}$.

Various theoretical and experimental studies have proposed mechanisms to realize the simultaneous breaking of $\mathcal{P}$ and $\mathcal{T}$, 
leading to the emergence of the JDE. 
In multiterminal Josephson junctions, the supercurrent is described as a multivariate function of the superconducting phase differences~\cite{gupta-nc14, chiles-nlt23}. 
When direct current biases (DC biases) constrain some of these phases, the resulting current can be viewed as a function of reduced variables, corresponding to an asymmetric slice of the free energy landscape in phase space~\cite{zhang-pra21}. 
Consequently, the effective current--phase relation becomes nonreciprocal, giving rise to the JDE. 
This indicates that nonreciprocity can arise from geometrical constraints in phase space rather than intrinsic symmetry breaking in the conventional sense.
In contrast to such geometrical mechanisms, nonreciprocity can also arise from explicit $\mathcal{T}$ breaking induced by magnetic fields. 
For instance, the Nb/Au/$\rm{NbSe_{2}}$ Josephson junction exhibits the JDE due to the combined effect of $\mathcal{P}$ breaking arising from structural asymmetry and $\mathcal{T}$ breaking associated with Abrikosov vortices~\cite{tian-apl123}. 
Moreover, superconducting quantum interference devices (SQUIDs) with artificially engineered asymmetric geometries and external magnetic flux have also been shown to exhibit the JDE~\cite{souto-prl129, ciaccia-prr5, greco-apl123, coraiola-acs18}. 

In addition to these mechanisms, nonreciprocity can also emerge from intrinsic spin-dependent interactions such as spin--orbit interaction and exchange fields. 
In particular, systems with Rashba spin--orbit interaction (RSOI) and exchange fields can break $\mathcal{P}$ and $\mathcal{T}$, respectively. 
When both symmetries are simultaneously broken, the JDE can emerge in such systems~\cite{chen-prb98, baumgartner-nt17, zhang-prx12, wei-prb108, kim-nc15, huang-apl125, fracassi-apl124, hikino-jpsj94}. 
In such systems, the propagation of Cooper pairs becomes spin-dependent due to the interplay between the spin--orbit interaction and the exchange field. 
In particular, the RSOI induces a difference in effective propagation wave vectors for different spin components, 
while the exchange field enables the coupling between these components. 
As a result, different spin components acquire different propagation phases, leading to spin-dependent phase accumulation. 
This mechanism plays a crucial role in generating nonreciprocal superconducting transport. 
Notably, in these systems, the $\mathcal{P}$ and $\mathcal{T}$ breaking originate from the intrinsic properties of the material itself, rather than from artificial structural engineering such as interfaces or heterostructures.

In contrast to charge transport, nonreciprocal effects in spin Josephson currents remain largely unexplored. 
In $s$-wave superconductor/ferromagnet (S/F) hybrid junctions, spin-triplet Cooper pairs can be induced due to the presence of exchange fields, enabling spin-polarized Josephson currents~\cite{bergeret-rmp, bergeret-prl86, volkov-prb90, bergeret-prb68, houzet-prb76, volkov-prb81, robinson-science, khaire}. 
These spin-triplet correlations provide a natural platform for spin Josephson transport. 
Hereafter, spin Josephson transport denotes the dissipationless flow of spin angular momentum mediated by spin-triplet superconducting correlations.
Despite this, the possibility of nonreciprocal spin transport, namely the spin Josephson diode effect (SJDE), has not been fully clarified. 

Recent studies have demonstrated that the SJDE can arise from noncoplanar magnetic configurations.
For example, in $s$-wave superconductor/ferromagnet hybrid junctions with multiple magnetic layers, a finite scalar spin chirality, characterized by $\boldsymbol{m}_1 \cdot (\boldsymbol{m}_2 \times \boldsymbol{m}_3)$, simultaneously breaks $\mathcal{P}$ and $\mathcal{T}$ symmetries.
As a result, both charge and spin Josephson diode effects can emerge~\cite{schulz-prb112,sun-prb112}.
In these systems, the SJDE originates from a geometrical phase acquired by Cooper pairs propagating through a noncoplanar magnetic texture.
This phase can be interpreted as a Berry-phase effect associated with the real-space spin chirality.

By contrast, the mechanism considered in the present work does not rely on noncoplanar magnetic structures. 
Instead, the SJDE emerges from the interplay between Rashba spin-orbit interaction and the exchange field, 
which generates a momentum-space effective magnetic field and a finite $\varphi_0$ phase shift. 
Therefore, the present mechanism provides an alternative route to nonreciprocal spin transport based on spin-orbit coupling rather than real-space magnetic chirality. 
An important consequence is that the SJDE can be realized without requiring multiple noncoplanar magnetic layers or a finite scalar spin chirality. 
This distinction may be advantageous for experimental implementations because the Rashba spin-orbit interaction 
and exchange field can be controlled independently through material design and magnetic engineering. 

In addition, the relationship between the SJDE and the conventional Josephson diode effect (JDE) in charge transport remains unclear. 
In conventional JDE systems, it has been shown that the $\varphi_{0}$ phase shift can be very small 
in the presence of a thin ferromagnetic layer~\cite{hikino-ph}. 
A sizable $\varphi_{0}$ shift can be achieved by introducing a sufficiently thick ferromagnetic layer 
that suppresses the spin-singlet component~\cite{hikino-jpsj94}. 
This suggests that the suppression of the singlet component plays an important role in enhancing nonreciprocal charge transport. 
In particular, it is not obvious whether the suppression of spin-singlet correlations is a necessary condition for realizing the SJDE. 
Clarifying this point is crucial for understanding whether the SJDE is simply a spin analogue of the JDE or represents a fundamentally distinct mechanism of nonreciprocal transport.

To address these issues, we theoretically study the SJDE in a diffusive Josephson junction incorporating a Rashba metal (RM) layer in the presence of a ferromagnetic exchange field. 
Within the quasiclassical Green's function framework, we derive analytical expressions for both the first- and second-harmonic spin Josephson currents (FHSJC and SHSJC). 
We show that these spin Josephson currents (SJCs) exhibit damped oscillatory behavior as a function of the RM thickness, accompanied by sign reversals of the SJC, reflecting the interference between different pairing correlations. 
A similar crossover behavior is also observed with increasing RSOI strength. 
Importantly, a finite $\varphi_{0}$ phase shift emerges in the spin current--phase relation of both harmonic components even in the absence of an external magnetic field. 
This intrinsic phase shift leads to the realization of the SJDE without requiring external fields. 
Furthermore, we demonstrate that the SJDE can be achieved without suppressing the spin-singlet component, in contrast to conventional JDE systems. 
Finally, we evaluate the efficiency of the SJDE within the perturbative regime.

\section{\texorpdfstring{S/F$_{\rm L}$/F/RM/F$_{\rm R}$/S junction}{S/FL/F/RM/FR/S junction}}
\begin{figure}
 \centering
        \includegraphics[width=0.5\textwidth]{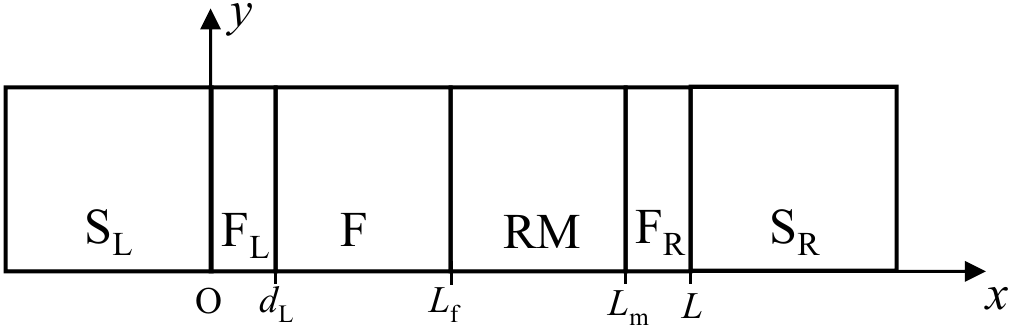}
 \caption{The schematic diagram of the ${\rm S}_{\rm L}/{\rm F}_{\rm L}/{\rm F}/{\rm RM}/{\rm F}_{\rm R}/{\rm S}_{\rm R}$ junction is shown, 
where ${\rm S}_{\rm L}$ (${\rm S}_{\rm R}$) denotes the spin-singlet superconductor on the left (right), 
${\rm F}_{\rm L}$ (${\rm F}_{\rm R}$) the thin ferromagnetic metal on the left (right), ${\rm F}$ the thick ferromagnetic metal, 
and normal metal with the RSOI (RM). 
The thick ferromagnet ${\rm F}$ serves as a barrier that suppresses the contribution of spin-singlet Cooper pairs. 
The total thicknesses of different regions are defined as $L_{\rm f} = d_{\rm L} + d_{\rm f}$, $L_{\rm m} = L_{\rm f} + d_{\rm m}$, and $L = L_{\rm m} + d_{\rm R}$, 
where $d_{\rm L(R)}$ are the thicknesses of ${\rm F}_{\rm L(R)}$, $d_{\rm f}$ that of ${\rm F}$, and $d_{\rm m}$ that of ${\rm RM}$.}
\label{fig1}
\end{figure}

\subsection{Nonlinear Usadel equation and harmonic decomposition}
As shown schematically in figure 1, we consider an $\rm{S/F_{\rm F}/F/RM/F_{\rm}/S}$ junction, 
where $\rm{S_{\rm L}}$ $(\rm{S_{\rm R}})$ denotes the spin-singlet superconductor on the left (right), 
$\rm{F_{\rm L}}$ ($\rm{F_{\rm R}}$) denotes the thin ferromagnetic layer on the left (right),
F is a thick ferromagnetic layer, and RM is a normal-metal layer with Rashba spin–orbit interaction (RSOI). 
The thick ferromagnetic layer acts as a barrier that suppresses the propagation of spin-singlet Cooper pairs.
We start from the full nonlinear Usadel equation and perform a perturbative expansion of the Green's function under the normalization condition\cite{richard-prl110}.
This procedure enables us to systematically derive the Usadel equations governing the first- and second-harmonic components of the spin Josephson current (SJC).

The Usadel equation in region $j$ ($={\rm F}_{\rm L}$, F, RM, and ${\rm F}_{\rm R}$) is given by
\begin{align}
-i \hbar D \tilde{\partial}_{x} 
\left[
\check {g}^{ j }(x) \tilde \partial _x \check {g}^{ j }(x)
\right]
+
\left[
\check{G}_{0}^{-1},\check {g}^{ j }(x)
\right]
&= \check{0},
\label{usadel1}
\end{align}
with the covariant derivative
\begin{eqnarray}
{{\tilde \partial }_x} = \left\{ \begin{array}{l}
{\partial _x} \bullet  - i{\alpha _{\rm R}}\left[ {{{\hat \sigma }_y}, \bullet } \right], \quad L_{\rm f} < x < L_{\rm m},\\
{\partial _x} \bullet , \quad \text{otherwise}.
\end{array} \right.
\end{eqnarray}
Here $D$ is the diffusion coefficient, which is assumed to be identical in all regions.
$\omega_{n} = (2n + 1)\pi k_{\rm B} T/\hbar$ is the fermion Matsubara frequency, and
$[\hat A, \hat B]$ denotes the commutator.
$\alpha_{\rm R}$ is the Rashba spin--orbit interaction (RSOI) strength in the RM,
and $\hat{\sigma}_{y}$ is the $y$ component of the Pauli matrices in spin space.

The quasiclassical Green's function $\check{g}^{j}(x)$ is expressed in particle--hole space as
\begin{equation}
\check{g}^{j}(x) =
\begin{pmatrix}
\hat{G}^{j}(x) & \hat{F}^{j}(x) \\
\hat{\bar{F}}^{j}(x) & \hat{\bar{G}}^{j}(x)
\end{pmatrix},
\label{gj-ph}
\end{equation}
where each matrix element is a $2\times2$ matrix in spin space.
The resolvent $\check{G}^{-1}_{0}$ is given by
\begin{equation}
\check{G}_{0}^{-1}
=
i\hbar \omega_{n} \check{\rho}_{z}
+
\check{\Sigma}_{\rm ex}\check{\rho}_{z},
\end{equation}
with $\check{\rho}_{z}$ being the $z$ component of the Pauli matrices in particle--hole space.
The exchange field contribution $\check{\Sigma}_{\rm ex}$ is defined as
\begin{equation}
\check{\Sigma}_{\rm ex}\check{\rho}_{z}
=
\begin{pmatrix}
\bm{h}(x)\cdot \hat{\bm{\sigma}} & \hat{0} \\
\hat{0} & \bm{h}(x)\cdot \hat{\bm{\sigma}}^{*}
\end{pmatrix},
\label{sigma-ex}
\end{equation}
where $\hat{\bm \sigma}=(\hat{\sigma}_{x}, \hat{\sigma}_{y}, \hat{\sigma}_{z})$.
The exchange field $\bm{h}(x)$ is modeled as
\begin{align}
\bm h(x) &= \left\{
\begin{array}{l}
h_{x}^{\rm L}\bm e_x + h_{z}^{\rm L}\bm e_z, \quad 0 < x < d_{\rm L},\\
h\,\bm e_z, \quad d_{\rm L} < x < L_{\rm f},\\
0, \quad L_{\rm f} < x < L_{\rm m},\\
h_{x}^{\rm R}\bm e_x + h_{z}^{\rm R}\bm e_z, \quad L_{\rm m} < x < L,
\end{array}
\right.
\label{exchange}
\end{align}
where $h_{x}^{\rm L(R)} = h \sin \theta_{\rm L(R)}$ and
$h_{z}^{\rm L(R)} = h \cos \theta_{\rm L(R)}$.

Applying the normalization condition $\check{g}(x)^2=\check{1}$,
the nonlinear Usadel equation corresponding to the $(1,2)$ component
in particle--hole space is written as
\begin{align}
-i\hbar D \tilde{\partial}_{x}
\Big[
\sqrt{1-\hat{F}^{j}\hat{\bar{F}}^{j}}\,\tilde{\partial}_{x} \hat{F}^{j}
-
\hat{F}^{j}\tilde{\partial}_{x}\sqrt{1-\hat{\bar{F}}^{j}\hat{F}^{j}}
\Big]
+i2\hbar \omega_{n} \hat{F}^{j}
+\bm{h}\cdot \hat{\bm{\sigma}} \hat{F}^{j}
-\hat{F}^{j}\bm{h}\cdot \hat{\bm{\sigma}}^{*}
= \hat{0}.
\label{usadel2}
\end{align}

Since we focus on the vicinity of the superconducting transition temperature $T_{\rm c}$,
the pair potential is small enough to satisfy $\Delta/\omega_n \ll 1$.
This justifies the perturbative expansion of the anomalous Green's functions 
and the linearization of the Usadel equation with respect to the small parameter $\Delta/\omega_n$.
On this basis, we derive the first- and second-harmonic SJCs analytically.
To solve equation~(\ref{usadel2}) perturbatively near $T_{\rm c}$, the anomalous Green's function is expanded as
\begin{equation}
\hat{F}^{j}(x) \approx \hat{f}^{j,(1)}(x) + \hat{f}^{j,(2)}(x)
\label{f-approx}
\end{equation}
Here, $\hat{f}^{j,(1)}$ and $\hat{f}^{j,(2)}$ represent the leading- and next-to-leading-order contributions
in the perturbative expansion with respect to $\Delta/\omega_n$.
The former gives rise to the first-harmonic spin Josephson current (FHSJC),
while the latter generates the second-harmonic component
through nonlinear coupling effects in the Usadel equation.
Since we focus on the vicinity of $T_{\rm c}$, the dominant contribution arises from the lowest Matsubara frequency.
Therefore, we retain only the $n=0$ component in the following analysis.
Substituting equation~(\ref{f-approx}) into equation~(\ref{usadel2})
and expanding $\sqrt{1-x}$ for small $x$,
we obtain the linearized Usadel equation for the first-order anomalous Green's function,
\begin{eqnarray}
\tilde{\partial}_{x}^{2} \hat{f}^{j,(1)}
-
\frac{2 \omega_{n}}{D} \hat{f}^{j,(1)}
+
i\frac{\bm{h}\cdot \hat{\bm{\sigma}}}{\hbar D} \hat{f}^{j,(1)}
-
i\hat{f}^{j,(1)}\frac{\bm{h}\cdot \hat{\bm{\sigma}}^{*}}{\hbar D}
=
\hat{0},
\label{usadel3}
\end{eqnarray}
and the linear Usadel equation for the second-order anomalous Green's function,
\begin{eqnarray}
\tilde{\partial}_{x}^{2} \hat{f}^{j,(2)}
-
\frac{2 \omega_{n}}{D} \hat{f}^{j,(2)}
+
i\frac{\bm{h}\cdot \hat{\bm{\sigma}}}{\hbar D} \hat{f}^{j,(2)}
-
i\hat{f}^{j,(2)}\frac{\bm{h}\cdot \hat{\bm{\sigma}}^{*}}{\hbar D}
\nonumber\\
=
\frac{1}{2}
\tilde{\partial}_{x}
\Big[
\hat{f}^{j,(1)} \hat{\bar{f}}^{j,(1)} \tilde{\partial}_{x} \hat{f}^{j,(1)}
-
\hat{f}^{j,(1)} \tilde{\partial}_{x}
\big(
\hat{\bar{f}}^{j,(1)} \hat{f}^{j,(1)}
\big)
\Big].
\label{usadel4}
\end{eqnarray}
It should be emphasized that the second-harmonic spin Josephson current (SHSJC) is not assumed a priori,
but is induced by the nonlinear coupling of the first-order anomalous Green's functions,
as described by equation~(\ref{usadel4}).

\subsection{Anomalous Green's functions contributing to the first-harmonic spin Josephson current}
In this subsection, we derive the anomalous Green's functions that contribute to the FHSJC. 
Here, we derive the anomalous Green's functions in the RM that contribute to a SJC with a $2\pi$-periodic phase dependence, 
since we focus on the contribution to the FHSJC flowing through the RM. 

Such an anomalous Green's function is expressed as 
\begin{eqnarray}
\hat{f}^{j,(1)}(x) = 
\left(
f^{j,(1)}_{s}(x) + \bm{f}^{j,(1)} (x)\cdot \hat{\bm{\sigma}}
\right)
i \hat{\sigma}_{y}
= \left( {\begin{array}{*{20}{c}}
{ - f_{x}^{j,(1)}\left( x \right) + if_{y}^{j,(1)}\left( x \right)}&{f_{s}^{j,(1)}\left( x \right) + f_{z}^{j,(1)}\left( x \right)}\\
{ - f_{s}^{j,(1)}\left( x \right) + f_{z}^{j,(1)}\left( x \right)}&{f_{x}^{j,(1)}\left( x \right) + if_{y}^{j,(1)}\left( x \right)}
\end{array}} \right).
\label{fj1}
\end{eqnarray}
where $f^{j,(1)}_{s}(x)$ represents the anomalous Green's function of the spin-singlet Cooper pairs (SSC), 
while $f^{j,(1)}_{x(y)}(x)$ and $f^{j,(1)}_{z}(x)$ correspond to those of the spin-triplet Cooper pairs (STC) 
with $|S_{z}|=1$ and $|S_{z}|=0$, respectively.  

To obtain particular solutions of equation~(\ref{usadel3}), 
we apply suitable boundary conditions to its general solutions as follows~\cite{demler-prb},
\begin{eqnarray}
\hat f^{\rm S_{\rm L}}(x=0) &=& \hat f^{\rm F_{\rm L}, (1)}(x=0), 
\label{bc1}\\
\hat f^{\rm F_{\rm L}, (1)}(x=d_{\rm L}) &=& \hat f^{\rm F, (1) }(x=d_{\rm L}), 
\label{bc2}\\
{\left. \frac{d}{dx} {\hat f}^{\rm F_{\rm L}, (1)} (x) \right|_{x=d_{\rm L}}} &=& {\left. \frac{d}{dx} {\hat f}^{\rm F, (1)} (x) \right|_{x=d_{\rm L}}}, 
\label{bc3}\\
{\hat f}^{\rm F, (1)} \left(x=L_{\rm f} \right) &=& {\hat f}^{\rm RM, (1)} \left(x=L_{\rm f} \right), 
\label{bc4} \\
{\left. \frac{d}{dx} {\hat f}^{\rm F, (1)} (x) \right|_{x=L_{\rm f}}} &=& {\left. \frac{d}{dx} {\hat f}^{\rm RM, (1)} (x) \right|_{x=L_{\rm f}}},
\label{bc5} \\
{\hat f}^{\rm RM, (1)} \left( x=L_{\rm m} \right) &=& {\hat f}^{\rm F_{\rm R}, (1)} \left( x=L_{\rm m} \right), 
\label{bc6} \\
{\left. \frac{d}{dx} {\hat f}^{\rm RM, (1)} (x) \right|_{x=L_{\rm m}}} &=& {\left. \frac{d}{dx} {\hat f}^{\rm F_{\rm R}, (1)} (x) \right|_{x=L_{\rm m}}}, 
\label{bc7} \\
\hat f^{\rm F_{\rm R}, (1)}(x=L) &=& \hat f^{\rm S_{\rm R}}(x=L). 
\label{bc8}
\end{eqnarray}
Starting from the anomalous Green's functions $\hat{f}^{\rm F_{\rm L (R)}, (1)}(x)$, 
we perform a Taylor expansion with respect to $x$ assuming $d_{\rm L (R)}/\xi \ll 1$, 
and apply equations~(\ref{bc1})--(\ref{bc3}) (equations~(\ref{bc6})--(\ref{bc8})) to $\hat{f}^{\rm F_{\rm L (R)}, (1)}(x)$.
As a result, $\hat{f}^{\rm F_{\rm L}, (1)}(x)$ and $\hat{f}^{\rm F_{\rm R}, (1)}(x)$ are obtained as 
%
\begin{eqnarray}
\hat f ^{\rm F_{\rm L},(1)} (d_{\rm L}) & \approx &
d_{\rm L} \partial _x { \hat f ^{\rm F,(1)}(d_{\rm L}) } + \hat f^{\rm S_{ \rm L } }(0)
+ i \frac{ d_{\rm L}^{2} h_{x}^{\rm L} }{ 2 \hbar D }  \left[ {{{\hat \sigma }_x},\hat f^{\rm S_{ \rm L } }(0) } \right] 
+ i \frac{ d_{\rm L}^{2} h_{z}^{\rm L} }{ 2 \hbar D }  \left[ {{{\hat \sigma }_z},\hat f^{\rm S_{ \rm L } }(0)} \right] 
\label{gf-fl}, 
\end{eqnarray}
and 
%
\begin{eqnarray}
\hat f ^{\rm F_{\rm R},(1)} (L_{\rm m}) & \approx &
-d_{\rm R} \partial _x { \hat f ^{\rm RM,(1)}(L_{\rm m}) } + \hat f^{\rm S_{ \rm R } }(L)
+ i \frac{ d_{\rm R}^{2} h_{x}^{\rm R} }{ 2 \hbar D }  \left[ {{{\hat \sigma }_x},\hat f^{\rm S_{ \rm R } }(L) } \right] 
+ i \frac{ d_{\rm R}^{2} h_{z}^{\rm R} }{ 2 \hbar D }  \left[ {{{\hat \sigma }_z},\hat f^{\rm S_{ \rm R } }(L)} \right] 
\label{gf-fr}. 
\end{eqnarray}
For details of the calculation, refer  to Refs~\cite{houzet-prb76} and \cite{hikino-prb}. 
Here, we adopt the anomalous Green's function of the $S_{\rm L (R)}$ near $T_{\rm c}$ within the rigid boundary condition. 
Therefore, the anomalous Green's function of the $S_{\rm L (R)}$ is given by
%
\begin{eqnarray}
\hat f^{\rm S_{\rm L ({\rm R})}}(x)|_{x=0(L)} 
= - \hat \sigma_{y}
\frac{ \Delta_{\rm L(R)} }{ \hbar \omega }
\label{fs}, 
\end{eqnarray}
where $\Delta_{\rm L(R)}$ is the superconducting gap of the ${\rm S}_{\rm L (R)}$. 

To obtain anomalous Green's functions in the RM, 
we solve the Usadel equation in the F and the RM. 
General solutions in the F and the RM are given by 
%
\begin{eqnarray}
\left( \begin{array}{l}
f_{s}^{{\rm{F}},(1)}\left( x \right)\\
f_{x}^{{\rm{F}},(1)}\left( x \right)\\
f_{z}^{{\rm{F}},(1)}\left( x \right)
\end{array} \right) &=& 
A \left( \begin{array}{l}
0\\
1\\
0
\end{array} \right){e^{ x/\xi} } + 
B \left( \begin{array}{l}
0\\
1\\
0
\end{array} \right){e^{ -x/\xi} } \nonumber \\
&+&
C \left( \begin{array}{l}
1\\
0\\
-1
\end{array} \right){e^{\kappa_{+} x} } + D \left( \begin{array}{l}
1\\
0\\
-1
\end{array} \right){e^{ -\kappa_{+} x}} 
 + E \left( \begin{array}{l}
1\\
0\\
1
\end{array} \right){e^{ \kappa_{-} x }} 
 + F\left( \begin{array}{l} 
1\\
0\\
1
\end{array} \right){e^{ -\kappa_{-} x }} 
\label{gs-f}, \nonumber \\
\end{eqnarray}
and 
%
\begin{eqnarray}
\left( \begin{array}{l}
f_s^{{\rm{RM}},(1)}\left( x \right)\\
f_{x}^{{\rm{RM}},(1)}\left( x \right)\\
f_{z}^{{\rm{RM}},(1)}\left( x \right)
\end{array} \right) &=& 
M \left( \begin{array}{l}
1\\
0\\
0
\end{array} \right){e^{ x/\xi }} + N \left( \begin{array}{l}
1\\
0\\
0
\end{array} \right){e^{ -x/\xi }} \nonumber \\
 &+& G\left( \begin{array}{l}
0\\
-i\\
i
\end{array} \right){e^{i 2\alpha_{\rm R} x}}{e^{ x/\xi }} + H\left( \begin{array}{l} 
0\\
-i\\
i
\end{array} \right){e^{i2\alpha_{\rm R} x}}{e^{ -x/\xi }} \nonumber \\
&+& I\left( \begin{array}{l}
0\\
i\\
i
\end{array} \right){e^{ - i2\alpha_{\rm R} x}}{e^{ x/\xi }} 
+ J\left( \begin{array}{l}
0\\
i\\
i
\end{array} \right){e^{ - i2\alpha_{\rm R} x}}{e^{ -x/\xi }} 
\label{gs-frm}. 
\end{eqnarray}
Where $\xi = \sqrt{\hbar D /2\pi k_{\rm B} T}$ and $\kappa_{\pm} = \sqrt{(\hbar \omega \pm i h)/\hbar D}$. 
It should be noted that $f_{ty}^{\rm F (RM)} (x)$ is exactly zero because the exchange field has no $y$ component~\cite{houzet-prb76,hikino-prb}.
To obtain special solutions of equation~(\ref{gs-frm}), we assume that $d_{\rm f}$ and $d_{\rm m}$ are much larger than $\xi$ 
for simplicity in the calculation of $\hat{f}^{\rm RM, (1)}(x)$. 
Applying equations~(\ref{bc4}) and (\ref{bc5}) to equations~(\ref{gs-f}) and (\ref{gs-frm}), and then substituting equations~(\ref{fs}), (\ref{gs-f}), 
and (\ref{gs-frm}) into equations~(\ref{gf-fl}) and (\ref{gf-fr}), we obtain the simultaneous algebraic equation to determine the coefficients 
satisfying the boundary conditions. 
By solving these simultaneous algebraic equations, the anomalous Green's functions in the RM can be approximately expressed as
%
%
\begin{equation}
f_{s}^{\rm RM,(1)}(x) \approx i \frac{\Delta_{\rm R}}{ \hbar \omega } e^{(x-L_{\rm m})/\xi}
\label{fs-rm},
\end{equation}
\begin{eqnarray}
f_{x}^{\rm RM,(1)}(x) &\approx&
	\frac{2 D_{x}^{\rm L}}{ 1+\kappa_{\alpha}^{+} \xi }
	\left[
	e^{- 2 d_{\rm m}/\xi} e^{\kappa_{\alpha}^{-} \left(x-d_{\rm f} \right) }
	-e^{- \kappa_{\alpha}^{+} \left(x-d_{\rm f} \right)}
	\right]
	e^{-d_{\rm f}/\xi} \frac{\Delta_{\rm L}}{ \hbar \omega } \nonumber \\
	&+&
	\left[
	\frac{\alpha_{\rm R}^{-}}{ 2 } \left(D_{x}^{\rm R} - D_{z}^{\rm R} \right) e^{-\kappa_{\alpha}^{-} \left(x-L_{\rm m} \right)} 
	-
	\frac{\alpha_{\rm R}^{+}}{ 2 } \left(D_{x}^{\rm R} + D_{z}^{\rm R} \right) e^{-\kappa_{\alpha}^{+} \left(x-L_{\rm m} \right)} 
	\right]
	\frac{ \Delta_{\rm R} }{ \hbar \omega } \nonumber \\
	&+&
	\frac{ {\Sigma}_{xz}^{\rm R} }{1+\kappa_{\alpha}^{+}\xi  }
	\left[
	e^{-\kappa_{\alpha}^{+} \left(x-d_{\rm f} \right)} - e^{\kappa_{\alpha}^{-} \left(x-d_{\rm f} \right)} e^{-2 d_{\rm m}/\xi} 
	\right]
	\frac{\Delta_{\rm R}}{\hbar \omega}
\label{fx-rm}, \\
 f_{z}^{\rm RM,(1)}(x) &\approx&
	\frac{2 D_{x}^{\rm L}}{ 1+\kappa_{\alpha}^{+} \xi }
	\left[
	e^{- 2 d_{\rm m}/\xi} e^{\kappa_{\alpha}^{-} \left(x-d_{\rm f} \right) }
	-e^{- \kappa_{\alpha}^{+} \left(x-d_{\rm f} \right)}
	\right]
	e^{-d_{\rm f}/\xi} \frac{\Delta_{\rm L}}{ \hbar \omega } \nonumber \\
	&-&
	\left[
	\frac{\alpha_{\rm R}^{+}}{ 2 } \left(D_{x}^{\rm R} + D_{z}^{\rm R} \right) e^{-\kappa_{\alpha}^{+} \left(x-L_{\rm m} \right)} 
	+
	\frac{\alpha_{\rm R}^{-}}{ 2 } \left(D_{x}^{\rm R} - D_{z}^{\rm R} \right) e^{-\kappa_{\alpha}^{-} \left(x-L_{\rm m} \right)} 
	\right]
	\frac{ \Delta_{\rm R} }{ \hbar \omega } \nonumber \\
	&+&
	\frac{ {\Sigma}_{xz}^{\rm R} }{1+\kappa_{\alpha}^{+}\xi  }
	\left[
	e^{-\kappa_{\alpha}^{+} \left(x-d_{\rm f} \right)} - e^{\kappa_{\alpha}^{-} \left(x-d_{\rm f} \right)} e^{-2 d_{\rm m}/\xi} 
	\right]
	\frac{\Delta_{\rm R}}{\hbar \omega}
\label{fz-rm},
\end{eqnarray}
and
\begin{eqnarray}
{\Sigma}_{xz}^{\rm R} = 
i D_{x}^{\rm R}{\rm Im}
\left[
\alpha_{\rm R}^{+} \left(1 + \kappa_{\alpha}^{+} \xi \right) e^{\kappa_{\alpha}^{+}d_{\rm m}}
\right]
+
D_{z}^{\rm R}{\rm Re}
\left[
\alpha_{\rm R}^{+} \left(1 + \kappa_{\alpha}^{+} \xi \right) e^{\kappa_{\alpha}^{+}d_{\rm m}}
\right]. 
\end{eqnarray}
Where $\kappa_{\alpha}^{\pm} = 1/\xi \pm i 2 \alpha_{\rm R}$, 
$D_{x (z)}^{\rm L (R)} = d_{\rm L (R)}^{2} h_{x (z)}^{\rm L (R)}/\hbar D$, and $\alpha_{\rm R}^{\pm}=1 \pm i 2\alpha_{\rm R}d_{\rm R}$. 
In the next subsection, we calculate the anomalous Green's function contributing to SHSJC by solving equation~(\ref{usadel4}). 

\subsection{Anomalous Green's function contributing to the second-harmonic spin Josephson current}
In this subsection, we derive the anomalous Green's functions that contribute to the SHSJC.
Unlike the first-harmonic case, the second-harmonic contribution arises from 
higher-order corrections to the anomalous Green's functions, without assuming the second harmonic a priori.
Here, we focus on the anomalous Green's functions in the RM that contribute to a SJC with a $4\pi$-periodic phase dependence,
since the second-harmonic contribution to the SHSJC originates from higher-order anomalous Green's functions.
The formal structure of the anomalous Green's functions $\hat{f}^{j,(2)}$ is identical to that of $\hat{f}^{j,(1)}$
and is therefore not repeated here.

To solve equation~(\ref{usadel4}), we impose homogeneous boundary conditions on the second-order correction, 
since no second-harmonic component is injected from the interfaces. 
If nonzero boundary conditions were imposed, artificial second-harmonic components would be injected from the interfaces, 
obscuring the intrinsic bulk origin of the SJDE.
As a result, only the particular solution contributes, while the homogeneous solution vanishes.

The anomalous Green's functions contributing to the SHSJC are obtained by solving equation~(\ref{usadel4}) 
and are proportional to $\left(\Delta/\hbar \omega\right)^{3}$.
Assuming that $\xi \alpha_{\rm R} \ll 1$ but $\alpha_{\rm R} \neq 0$ and $d_{\rm f(m)}/\xi \gg 1$, 
we substitute equations~(\ref{fs-rm})--(\ref{fz-rm}) into equation~(\ref{usadel4}) and subsequently integrate equation~(\ref{usadel4}) with respect to $x$.
As a result, the anomalous Green's functions are obtained as follows,
%
\begin{eqnarray}
f_{x}^{\rm RM, (2)}(x) & \approx &
	-\frac{D_{x}^{\rm L}}{1 + \kappa_{\alpha}^{-} \xi}
	\left[
	\frac{1}{8} \left(6 + i 7 \xi \alpha_{\rm R} \right) 
	e^{-2 d_{\rm m} / \xi} e^{\kappa_{\alpha}^{+} \left(x-d_{\rm f} \right) } 
	-
	i \frac{1}{2 \xi \alpha_{\rm R}} \left( 1- i \xi \alpha_{\rm R} \right) e^{-\kappa_{\alpha}^{-} \left(x-d_{\rm f} \right) } 
	\right] \nonumber \\
	&\times&
	e^{-d_{\rm f}/\xi} e^{2\left(x-L_{\rm m} \right)/\xi} 
	\left(
	\frac{\Delta_{\rm R}}{\hbar \omega}
	\right)^{2}
	\frac{\Delta_{\rm L}^{*}}{\hbar \omega}
\label{fx2}, 
\end{eqnarray}
and 
\begin{eqnarray}
f_{z}^{\rm RM, (2)}(x) & \approx &
	i\frac{D_{x}^{\rm L}}{1 + \kappa_{\alpha}^{-} \xi}
	\left[
	\frac{1}{4} \left(1 + i \right) 
	e^{-2 d_{\rm m} / \xi} e^{\kappa_{\alpha}^{+} \left(x-d_{\rm f} \right) } 
	+
	i \frac{1}{2 \xi \alpha_{\rm R}} \left( 1- i 2 \right) e^{-\kappa_{\alpha}^{-} \left(x-d_{\rm f} \right) } 
	\right] \nonumber \\
	&\times&
	e^{-d_{\rm f}/\xi} e^{2\left(x-L_{\rm m} \right)/\xi} 
	\left(
	\frac{\Delta_{\rm R}}{\hbar \omega}
	\right)^{2}
	\frac{\Delta_{\rm L}^{*}}{\hbar \omega}
\label{fz2}.
\end{eqnarray}
Within the present approximation, $f_{s}^{\rm RM, (2)}(x)$ and $f_{y}^{\rm RM, (2)}(x)$ vanish.
Therefore, only the spin-triplet components acquire finite values in the anomalous Green's functions contributing to the SHSJC.

%
\section{Formulation of first- and second-harmonic spin Josephson currents}
We adopt the $\rm{SU(2)}$ covariant formulation of the spin current,
which provides a consistent definition in the presence of spin--orbit interaction.
This formulation ensures gauge covariance and is widely used in quasiclassical theories of spin transport.

Our starting point for the spin current is given by~\cite{tokatly-prb96},
\begin{eqnarray}
\bm{J}^{a}(x) &=& 
-i \frac{\pi N_{\rm F}D }{8 \beta}
\sum_{i \omega_{n}} {\rm Tr}
\left[
\check{\sigma}^{a} \check{\rho}_{z} \check{g}^{j}(x) 
\tilde{\partial}_{x} \check{g}^{j}(x)
\right],
\label{js1} \\
\check{\sigma}^{a} &=&
\begin{pmatrix}
\hat{\sigma}^{a} & \hat{0} \\
\hat{0} & \hat{\sigma}^{a,T}
\end{pmatrix},
\label{sgm-a}
\end{eqnarray}
Here, $\tilde{\partial}_{x}$ denotes the SU(2) covariant derivative, which includes the spin--orbit interaction through the gauge field.
Note that, in the presence of spin--orbit interaction, the spin current is generally not conserved 
within the SU(2) covariant framework.
Here, $N_{\rm F}$ is the density of states at the Fermi level and $\beta=1/k_{\rm B}T$.
$\hat{\sigma}^{a}$ is the Pauli matrix for the spin component $a$ ($a=x,y,z$), and $\hat{\sigma}^{a,T}$ denotes its transpose.
${\rm Tr}[\cdots]$ represents the trace in the Nambu space.
We impose the normalization condition $\check{g}^{j}(x)\check{g}^{j}(x)=\check{\rho}_{0}$.
Near $T_{\rm c}$, the anomalous component is small, $|\hat{f}|\ll1$,
so that the diagonal (normal) component of $\check{g}^{j}(x)$ can be approximated by its bulk value.
Therefore, substituting equation~(\ref{f-approx}), the linearized spin currents are given by
%
\begin{eqnarray}
J^{a,(1)}(x) &=& 
-i \frac{\pi N_{\rm F}D }{8 \beta}
\sum_{i \omega_{n}} {\rm tr}
\left[
\hat{\sigma}^{a} \hat{\bar{f}}^{j,(1)}(x) \tilde{\partial}_{x} \hat{f}^{j,(1)}(x)
-
\hat{\sigma}^{a} \hat{f}^{j,(1)}(x) \tilde{\partial}_{x} \hat{\bar{f}}^{j,(1)}(x)
\right],
\label{ja1} \\ 
J^{a,(2)}(x) &=&
-i \frac{\pi N_{\rm F}D }{8 \beta}
\sum_{i \omega_{n}} {\rm tr}
\left[
\hat{\sigma}^{a} \hat{\bar{f}}^{j,(1)}(x) \tilde{\partial}_{x} \hat{f}^{j,(2)}(x)
+
\hat{\sigma}^{a} \hat{\bar{f}}^{j,(2)}(x) \tilde{\partial}_{x} \hat{f}^{j,(1)}(x)
\right.
\nonumber\\
&&\left.
-
\hat{\sigma}^{a} \hat{f}^{j,(1)}(x) \tilde{\partial}_{x} \hat{\bar{f}}^{j,(2)}(x)
-
\hat{\sigma}^{a} \hat{f}^{j,(2)}(x) \tilde{\partial}_{x} \hat{\bar{f}}^{j,(1)}(x)
\right],
\label{ja2}
\end{eqnarray}
where $j^{a,(1)}(x)$ and $j^{a,(2)}(x)$ denote the FHSJC and SHSJC, respectively,
and ${\rm tr}[\cdots]$ represents the trace in the spin space. 
Substituting equation~(\ref{fj1}) into equation~(\ref{ja1}), the FHSJCs are expressed as
%
\begin{eqnarray}
J^{x,(1)}(x) &=&
	\frac{\pi N_{\rm F} D}{2\beta}
	\sum_{i\omega_{n}}
	\left\{
	{\rm Re}
	\left[
	{\bar{f}}^{j,(1)}_{z}(x) \partial_{x} {f}_{y}^{j,(1)}(x)
	-
	{\bar{f}}^{j,(1)}_{y}(x) \partial_{x} {f}_{z}^{j,(1)}(x)
	\right]
	\right.
\nonumber\\
	&+&\left.
	{\rm Im}
	\left[
	{\bar{f}}_{x}^{j,(1)}(x) \partial_{x} {f}_{s}^{j,(1)}(x)
	-
	{\bar{f}}_{s}^{j,(1)}(x) \partial_{x} {f}_{x}^{j,(1)}(x)
	\right]
	\right\},
\label{jx1} \\
J^{y,(1)}(x) &=&
	-i \frac{\pi N_{\rm F} D}{2\beta}
	\sum_{i\omega_{n}}
	\left\{
	{\rm Re}
	\left[
	{\bar{f}}^{j,(1)}_{s}(x) \partial_{x} {f}_{y}^{j,(1)}(x)
	-
	{\bar{f}}^{j,(1)}_{y}(x) \partial_{x} {f}_{s}^{j,(1)}(x)
	\right]
	\right.
\nonumber\\
	&+&\left.
	{\rm Im}
	\left[
	{\bar{f}}_{z}^{j,(1)}(x) \partial_{x} {f}_{x}^{j,(1)}(x)
	-
	{\bar{f}}_{x}^{j,(1)}(x) \partial_{x} {f}_{z}^{j,(1)}(x)
	\right]
	\right\},
\label{jy1} \\
J^{z,(1)}(x) &=&
	\frac{\pi N_{\rm F} D}{2\beta}
	\sum_{i\omega_{n}}
	\left\{
	{\rm Re}
	\left[
	{\bar{f}}^{j,(1)}_{x}(x) \partial_{x} {f}_{y}^{j,(1)}(x)
	-
	{\bar{f}}^{j,(1)}_{y}(x) \partial_{x} {f}_{x}^{j,(1)}(x)
	\right]
	\right.
\nonumber\\
	&+&\left.
	{\rm Im}
	\left[
	{\bar{f}}_{z}^{j,(1)}(x) \partial_{x} {f}_{s}^{j,(1)}(x)
	-
	{\bar{f}}_{s}^{j,(1)}(x) \partial_{x} {f}_{z}^{j,(1)}(x)
	\right]
	\right\}.
\label{jz1}
\end{eqnarray}
Using $\hat{f}^{j,(n)}(x)=\left[f^{j,(n)}_{s}(x)+\bm{f}^{j,(n)}(x)\cdot\hat{\bm{\sigma}}\right] i\hat{\sigma}_{y}$ $(n=1,2)$,
the SHSJCs are obtained as
%
\begin{eqnarray}
J^{x,(2)}(x) &=&
	\frac{\pi N_{\rm F} D}{2\beta}
	\sum_{i\omega_{n}}
	\left\{
	{\rm Re}
	\left[
	{\bar f}_{z}^{j,(1)}(x) \partial_{x} f_{y}^{j,(2)}
	+
	{\bar f}_{z}^{j,(2)}(x) \partial_{x} f_{y}^{j,(1)}
	\right.
	\right.
\nonumber\\
	&-&
	\left.
	\left.
	{\bar f}_{y}^{j,(1)}(x) \partial_{x} f_{z}^{j,(2)}
	-
	{\bar f}_{y}^{j,(2)}(x) \partial_{x} f_{z}^{j,(1)}
	\right]
	\right.
\nonumber\\
	&+&
	\left.
	{\rm Im}
	\left[
	{\bar f}_{x}^{j,(1)}(x) \partial_{x} f_{s}^{j,(2)}
	+
	{\bar f}_{x}^{j,(2)}(x) \partial_{x} f_{s}^{j,(1)}
	-
	{\bar f}_{s}^{j,(1)}(x) \partial_{x} f_{x}^{j,(2)}
	-
	{\bar f}_{s}^{j,(2)}(x) \partial_{x} f_{x}^{j,(1)}
	\right]
	\right\},
\label{jx2} \\
J^{y,(2)}(x) &=&
	-i\frac{\pi N_{\rm F} D}{2\beta}
	\sum_{i\omega_{n}}
	\left\{
	{\rm Re}
	\left[
	{\bar f}_{y}^{j,(1)}(x) \partial_{x} f_{s}^{j,(2)}
	+
	{\bar f}_{y}^{j,(2)}(x) \partial_{x} f_{s}^{j,(1)}
	\right.
	\right.
\nonumber\\
	&-&
	\left.
	\left.
	{\bar f}_{s}^{j,(1)}(x) \partial_{x} f_{y}^{j,(2)}
	-
	{\bar f}_{s}^{j,(2)}(x) \partial_{x} f_{y}^{j,(1)}
	\right]
	\right.
\nonumber\\
	&+&
	\left.
	{\rm Im}
	\left[
	{\bar f}_{x}^{j,(1)}(x) \partial_{x} f_{z}^{j,(2)}
	+
	{\bar f}_{x}^{j,(2)}(x) \partial_{x} f_{z}^{j,(1)}
	-
	{\bar f}_{z}^{j,(1)}(x) \partial_{x} f_{x}^{j,(2)}
	-
	{\bar f}_{z}^{j,(2)}(x) \partial_{x} f_{x}^{j,(1)}
	\right]
	\right\},
\label{jy2} \\
J^{z,(2)}(x) &=&
	\frac{\pi N_{\rm F} D}{2\beta}
	\sum_{i\omega_{n}}
	\left\{
	{\rm Re}
	\left[
	{\bar f}_{x}^{j,(1)}(x) \partial_{x} f_{y}^{j,(2)}
	+
	{\bar f}_{x}^{j,(2)}(x) \partial_{x} f_{y}^{j,(1)}
	\right.
	\right.
\nonumber\\
	&-&
	\left.
	\left.
	{\bar f}_{y}^{j,(1)}(x) \partial_{x} f_{x}^{j,(2)}
	-
	{\bar f}_{y}^{j,(2)}(x) \partial_{x} f_{x}^{j,(1)}
	\right]
	\right.
\nonumber\\
	&+&
	\left.
	{\rm Im}
	\left[
	{\bar f}_{z}^{j,(1)}(x) \partial_{x} f_{s}^{j,(2)}
	+
	{\bar f}_{z}^{j,(2)}(x) \partial_{x} f_{s}^{j,(1)}
	-
	{\bar f}_{s}^{j,(1)}(x) \partial_{x} f_{z}^{j,(2)}
	-
	{\bar f}_{s}^{j,(2)}(x) \partial_{x} f_{z}^{j,(1)}
	\right]
	\right\},
\label{jz2}
\end{eqnarray}
It should be noted that, in the present formulation, the $y$-component carries an overall factor of $-i$.
Nevertheless, the physical spin current is real. 
The origin of this structure will be discussed later. 

In the following, we evaluate the FHSJC and SHSJC using the solutions of the Usadel equations.
Substituting equations~(\ref{fs-rm})--(\ref{fz-rm}) into equation~(\ref{ja1}),
the FHSJC is obtained as
%
\begin{eqnarray}
J^{y,(1)}(d_{\rm m}) &=&
	i\frac{8 \pi N_{\rm F}D}{\beta \xi} D_{x}^{\rm L} D_{x}^{\rm R}
	\left(
	\frac{\Delta}{\hbar \omega}
	\right)^{2}
\nonumber\\
	&\times&
	\left[
	\sin \left(\theta + \varphi_{0} \right)
	-\xi \alpha_{\rm R} \cos \left( \theta + \varphi_{0} \right)
	\right]
	e^{-(d_{\rm f}+d_{\rm m})/\xi},
\label{jjy1}
\end{eqnarray}
and substituting equations~(\ref{fs-rm})--(\ref{fz-rm}) and (\ref{fx2})--(\ref{fz2}) into equation~(\ref{ja2}),
the SHSJC is given by
%
\begin{eqnarray}
J^{y,(2)}(d_{\rm m}) &=&
	i\frac{8 \pi N_{\rm F}D}{\beta \xi}
	\frac{(D_{x}^{\rm L})^{2}}{\xi \alpha_{\rm R}}
	\left(
	\frac{\Delta}{\hbar \omega}
	\right)^{4}
\nonumber\\
	&\times&
	\left[
	\sin \left(2\theta + 2\varphi_{0} \right)
	-3 \cos \left( 2\theta + 2\varphi_{0} \right)
	\right]
	e^{-2(d_{\rm f}+d_{\rm m})/\xi},
\label{jjy2}
\end{eqnarray}
where $\varphi_{0}=2\alpha_{\rm R}d_{\rm m}$ is the phase shift in the spin current phase relation.
From equations~(\ref{jjy1}) and (\ref{jjy2}), it is clearly seen that both the first- and second-harmonic SJCs 
exhibit $\varphi_{0}$-shifted spin current--phase relations and contain additional cosine terms, even in the absence of an external magnetic field.
Consequently, the coexistence of the $\varphi_{0}$-shifted first- and second-harmonic SJCs 
naturally gives rise to the SJDE without requiring an external magnetic field. 

The emergence of the $\varphi_{0}$ phase shift can be understood from symmetry considerations.
In the present system, the RSOI breaks inversion symmetry, while the exchange field breaks time-reversal symmetry.
Only when both symmetries are simultaneously broken, additional cosine terms are allowed in the spin current--phase relation, and a finite $\varphi_{0}$ shift can appear.
This symmetry-based mechanism explains 
why the coexistence of the first- and second-harmonic components results in the SJDE even in the absence of an external magnetic field.

Finally, we justify why we focus on the $y$-component of the SJC. 
With $\Delta_{\rm L}=\Delta$ and $\Delta_{\rm R}=\Delta e^{i\theta}$,
the first-order anomalous Green's functions scale as
$\hat{f}^{(1)}\propto \Delta_{\rm L}$ and/or $\Delta_{\rm R}\propto e^{i\theta}$,
whereas the second-order corrections obtained from the inhomogeneous equations
scale as $\hat{f}^{(2)}\propto (\Delta_{\rm R})^{2}\Delta_{\rm L}^{*}\propto e^{i2\theta}$.
Substituting these phase factors into the spin current expressions,
the $x$- and $z$-components reduce to the first-harmonic dependence,
$J^{x(z)}\propto \sin\theta$ (and/or $\cos\theta$). 
The second-order terms always combine with first-order contributions, 
yielding only $\pm\theta$ phase factors.
In contrast, the $y$-component contains finite terms oscillating as
$\sin 2\theta$ (and/or $\cos 2\theta$).
Therefore, we focus on the $y$-component of the SJC, 
since the second-harmonic contribution is of crucial importance for the SJDE.

These results indicate that the SJC intrinsically involves coupling between different spin components.
This structure suggests that the SJC originates from the interplay between distinct spin sectors, rather than from a single pairing channel.
The physical implications of this feature, including its relevance to the realization of the SJDE, will be discussed in more detail later.

Before presenting the numerical results, we briefly comment on the physical meaning of the spin current in the presence of spin--orbit interaction.
It should be noted that spin current is not generally a conserved quantity 
in the presence of spin--orbit interaction because spin angular momentum can be exchanged with orbital degrees of freedom. 
Nevertheless, the local spin current remains a useful and physically meaningful quantity for characterizing spin transport phenomena. 
Indeed, spin-current-based descriptions are widely employed in the study of spin Hall effects, spin pumping, 
and related spintronic phenomena~\cite{rmp-77, rmp-87}. 
In the present work, we therefore use the local spin current as a probe of the underlying 
transport mechanism responsible for the SJDE.

\section{Numerical results}
In this section, we present numerical results for the SJC and evaluate the efficiency of the SJDE 
based on equations~(\ref{jjy1}) and (\ref{jjy2}).
In all numerical calculations, the parameters are fixed as
$d_{\rm L(R)}/\xi_{\rm f}=0.4$, $d_{\rm f}/\xi=3$, $\theta_{\rm L(R)}=\pi/4$,
and $T/T_{\rm c}=0.9$,
where $\theta_{\rm L(R)}$ denotes the polar angle of the magnetization in
${\rm F}_{\rm L(R)}$.
Here, $\xi_{\rm f}=\sqrt{\hbar D/h}$ is the characteristic penetration length
of Cooper pairs in the ferromagnetic region.
The coherence length in the RM is given by
$\xi=\sqrt{\hbar D/(2\pi k_{\rm B}T)}$,
which characterizes the penetration of Cooper pairs into the RM.
In the following calculations, the thickness of the RM is measured in units of $\xi$.
Moreover, since we focus on the vicinity of $T_{\rm c}$,
the temperature dependence of the superconducting gap is taken as
$\Delta=\Delta_{0}\sqrt{1-T/T_{\rm c}}$.
In all figures, the explicit dependence on the RM thickness $L_{\rm m}$
is omitted for simplicity.
While the analytical expressions derived in the previous section
clarify the origin of the first- and second-harmonic SJCs, 
numerical calculations are essential to visualize their relative magnitudes
and parameter dependence.
Based on these settings, we discuss the qualitative behavior of
the first- and second-harmonic SJCs below.

\subsection{Numerical results of first- and second-harmonic spin Josephson current} 
\begin{figure}[!h]
 \centering
        \includegraphics[width=1\textwidth]{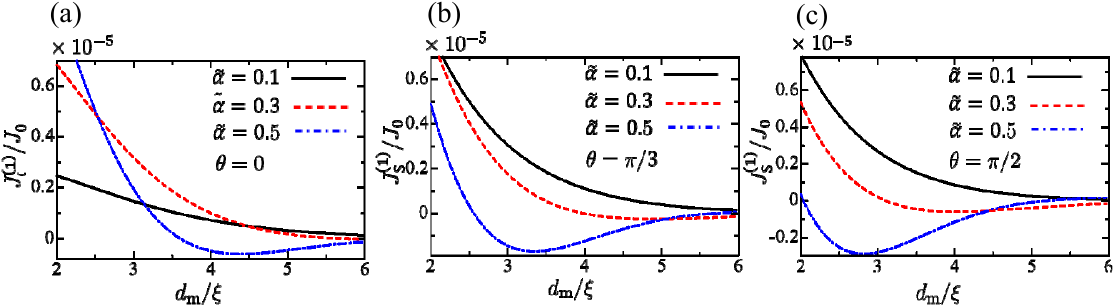}
\caption{
First-harmonic spin Josephson current (FHSJC) as a function of the thickness of the Rashba metal ($d_{\rm m}$) for $\tilde{\alpha}=\xi\alpha_{\rm R}=0.1$, $0.3$, and $0.5$.
Panels (a)--(c) correspond to $\theta=0$, $\pi/3$, and $\pi/2$, respectively.
For weak Rashba spin-orbit interaction ($\tilde{\alpha}=0.1$), the current exhibits a monotonic decay with increasing $d_{\rm m}$.
As $\tilde{\alpha}$ increases, a damped oscillatory behavior develops, and even for $\tilde{\alpha}=0.3$, a sign change can be identified, 
although the oscillation amplitude is small.
For $\tilde{\alpha}=0.5$, the oscillatory behavior becomes clearly visible for all values of $\theta$. 
The oscillatory behavior originates from the interplay between the Rashba spin-orbit interaction and the exchange field. 
The FHSJC is normalized by $J_{0}=i8\pi N_{\rm F}Dk_{\rm B}T_{\rm c}/\xi_{\rm c}$, where $\xi_{\rm c}=\sqrt{\hbar D/(2\pi k_{\rm B}T_{\rm c})}$.
}
\label{fig2}
\end{figure}
Figure~2 shows the $d_{\rm m}$ dependence of the FHSJC,
$J_{\rm s}^{(1)}=J^{y,(1)}(d_{\rm m})$, for $\theta=0$, $\pi/3$, and $\pi/2$.
The horizontal axis represents the thickness of the RM ($d_{\rm m}$), while the vertical axis denotes the FHSJC.
For weak RSOI ($\tilde{\alpha}=\alpha_{\rm R} \xi=0.1$), the FHSJC exhibits a nearly monotonic decay with increasing $d_{\rm m}$.
As $\tilde{\alpha}$ increases, a damped oscillatory behavior gradually develops, and even for $\tilde{\alpha}=0.3$, a sign change can be identified although the oscillation amplitude is small.
For $\tilde{\alpha}=0.5$, the oscillatory behavior becomes clearly visible for all values of $\theta$.
These results indicate that the oscillatory nature of the FHSJC is enhanced by the RSOI, and the direction of the FHSJC can be reversed by tuning the RM thickness $d_{\rm m}$. 
The damped oscillatory behavior can be understood in terms of the difference 
in effective propagation wave vectors for different spin components,
arising from the RSOI-induced difference in effective propagation wave vectors for different spin components,
in the presence of the exchange field. 

\begin{figure}[!h]
 \centering
        \includegraphics[width=1\textwidth]{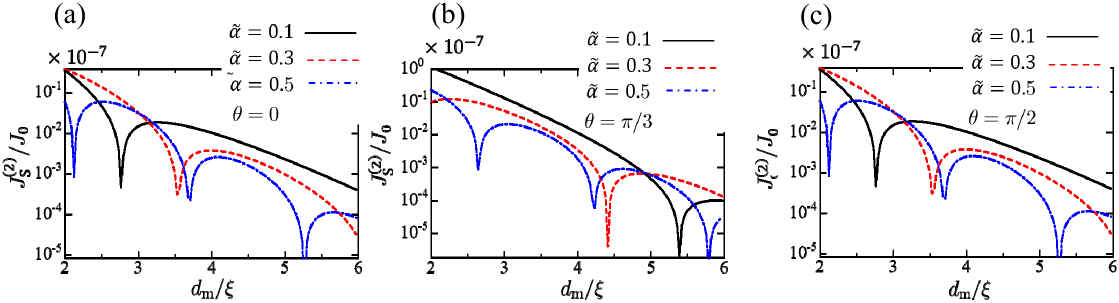}
 \caption{
Second-harmonic spin Josephson current (SHSJC) as a function of the thickness of the Rashba metal ($d_{\rm m}$) for $\tilde{\alpha}=\xi\alpha_{\rm R}=0.1$, $0.3$, and $0.5$.
Panels (a)--(c) correspond to $\theta=0$, $\pi/3$, and $\pi/2$, respectively.
The SHSJC is plotted on a logarithmic scale in terms of its absolute value.
For all values of $\tilde{\alpha}$, the SHSJC exhibits a damped oscillatory behavior as a function of $d_{\rm m}$. 
The logarithmic representation highlights the nodes associated with sign reversals as sharp dips.
The oscillation period becomes shorter with increasing $\tilde{\alpha}$, while the magnitude of the SHSJC is rapidly suppressed. 
The enhanced oscillatory behavior of the SHSJC reflects the higher-order coherent processes induced by the combined effects of the Rashba spin-orbit interaction and the exchange field.
The SHSJC is normalized by $J_{0}=i8\pi N_{\rm F}Dk_{\rm B}T_{\rm c}/\xi_{\rm c}$, where $\xi_{\rm c}=\sqrt{\hbar D/(2\pi k_{\rm B}T_{\rm c})}$.
}
\label{fig3}
\end{figure}
Figure~\ref{fig3} demonstrates the $d_{\rm m}$ dependence of the SHSJC,
$J_{\rm s}^{(2)}=\left|J^{y,(2)}(d_{\rm m})\right|$, 
for $\theta=0$, $\pi/3$, and $\pi/2$.
The horizontal axis represents the thickness of the RM ($d_{\rm m}$), while the vertical axis denotes the SHSJC.
Note that the vertical axis shows the absolute value of the SHSJC and is plotted on a logarithmic scale,
since $J_{\rm s}^{(2)}$ rapidly decreases with increasing $d_{\rm m}$.
The logarithmic representation also makes the nodes associated with sign reversals visible as sharp dips.
For all values of $\theta$, the SHSJC exhibits a pronounced damped oscillatory behavior as a function of $d_{\rm m}$.
Moreover, the oscillation period becomes shorter with increasing $\tilde{\alpha}$,
indicating an enhancement of the spin-dependent phase accumulation induced by the RSOI. 
In addition, the magnitude of the SHSJC decreases with increasing $\tilde{\alpha}$.
This behavior is consistent with equation~(42), which shows that $J^{y,(2)}(d_{\rm m})$ is inversely proportional to $\tilde{\alpha}$.
This inverse dependence indicates that, although the RSOI enhances the phase accumulation,
it simultaneously suppresses the amplitude of the second-harmonic component. 
These features demonstrate that the SHSJC is more sensitive to the spin-dependent phase accumulation than the FHSJC,
and plays a crucial role in generating nonreciprocal spin Josephson transport.

\begin{figure}[!h]
 \centering
        \includegraphics[width=0.7\textwidth]{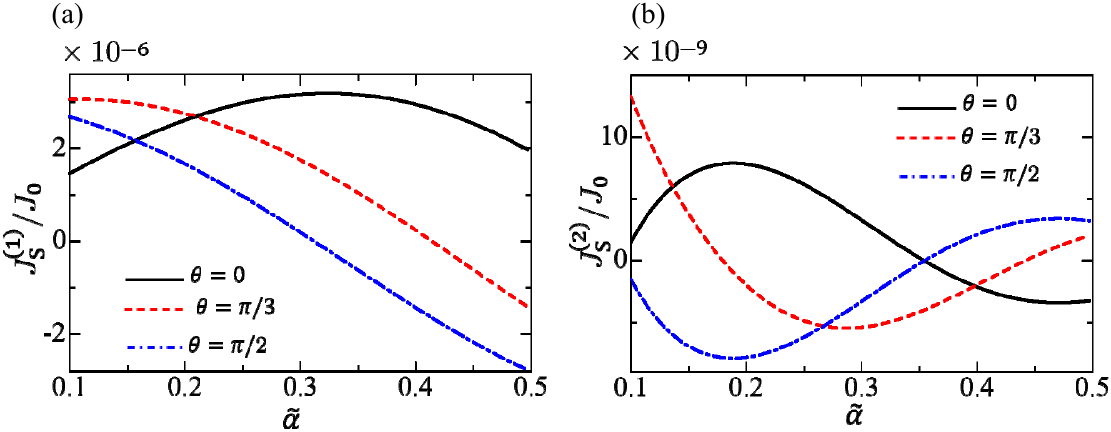}
 \caption{
The first- and second-harmonic spin Josephson currents,
$J_{\rm s}^{(1)}$ and $J_{\rm s}^{(2)}$,
as functions of the dimensionless Rashba spin--orbit interaction strength
${\tilde \alpha}$ for $\theta=0$, $\pi/3$, and $\pi/2$.
Panels (a) and (b) show the first- and second-harmonic spin Josephson currents,
respectively.
Both components change sign as ${\tilde \alpha}$ is varied,
while the second-harmonic component exhibits a more pronounced dependence. 
The pronounced ${\tilde \alpha}$ dependence of the SHSJC indicates that the Rashba spin--orbit interaction plays a key role in determining the efficiency of the spin Josephson diode effect.
The SHSJC is normalized by
$J_{0}=i8\pi N_{\rm F}Dk_{\rm B}T_{\rm c}/\xi_{\rm c}$,
where $\xi_{\rm c}=\sqrt{\hbar D/(2\pi k_{\rm B}T_{\rm c})}$.
}
\label{fig4}
\end{figure}
Figure~\ref{fig4} shows the dependence of the first- and second-harmonic SJCs on ${\tilde \alpha}$.
It is clearly seen that both components change sign as ${\tilde \alpha}$ is varied.
In particular, the second-harmonic component exhibits a more pronounced dependence on ${\tilde \alpha}$ compared to the first-harmonic component.
While the RSOI induces a phase shift and enhances the oscillatory behavior of the SJC,
the magnitude of the second-harmonic component is suppressed with increasing ${\tilde \alpha}$,
as expected from the inverse scaling $J^{y,(2)}(d_{\rm m}) \propto 1/{\tilde \alpha}$.
As a result, the relative contribution of the higher-order harmonic to the total SJC is reduced for large ${\tilde \alpha}$.
Consequently, although the RSOI modifies the spin current--phase relation,
the asymmetry of the current is not necessarily enhanced at large ${\tilde \alpha}$.
This tendency is reflected in the behavior of the SJDE efficiency, as shown in Figs.~\ref{fig7} and \ref{fig8}.
These results indicate that the nonreciprocal SJC is governed by a competition 
between the phase shift induced by the RSOI and the suppression of the second-harmonic component.

\begin{figure}[!h]
 \centering
        \includegraphics[width=0.45\textwidth]{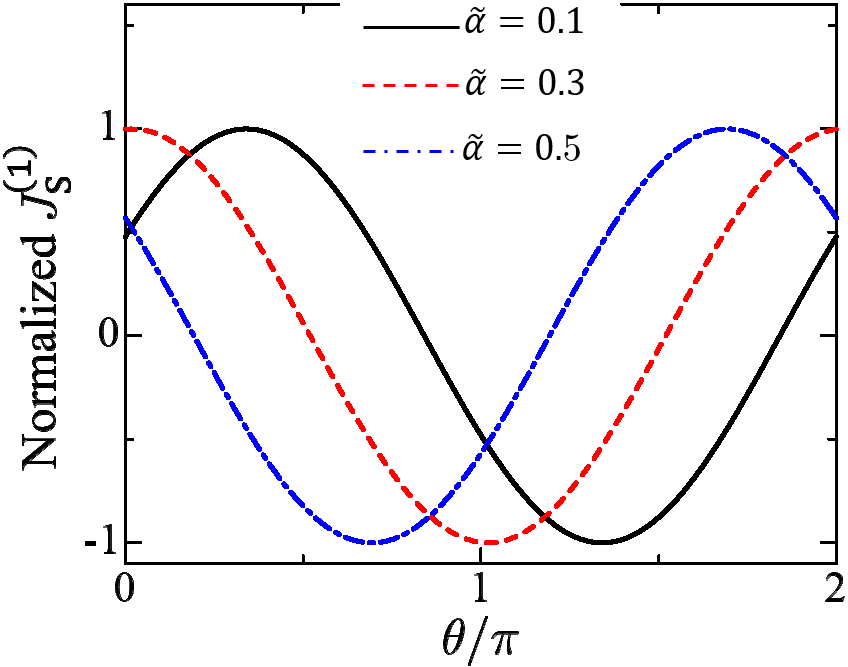}\caption{
The spin current--phase relation (SCPR) of the first-harmonic spin Josephson current (FHSJC) for $\tilde{\alpha}=0.1$, $0.3$, and $0.5$.
The FHSJC is normalized by its maximum value.
A sizable phase shift appears in the SCPR depending on $\tilde{\alpha}$.
The $\varphi_0$-phase shift provides a necessary ingredient for the SJDE, although an additional second-harmonic contribution is required to generate nonreciprocity.
}
\label{fig5}
\end{figure}
Figure~\ref{fig5} shows the spin current--phase relation (SCPR) for ${\tilde \alpha}=0.1$, $0.3$, and $0.5$,
which is a key result in determining whether the SJDE can be observed. 
The vertical axis represents the FHSJC normalized such that its maximum value is unity, 
while the horizontal axis denotes the superconducting phase difference between the two superconductors. 
In conventional Josephson junctions, the SCPR exhibits a simple sine function. 
This behavior is also recovered from Eq.~(\ref{jjy1}) by setting
$\alpha_{\rm R}=0$, yielding $J^{y,(1)}(d_{\rm m}) \propto \sin \theta$.
In contrast, in the present Josephson junction, the SCPR exhibits a phase-shifted form consisting of both sine 
and cosine components depending on ${\tilde \alpha}$,
namely,
\[
J^{y,(1)}(d_{\rm m})
\propto
\sin (\theta + \varphi_{0})
-
\xi \alpha_{\rm R} \cos (\theta + \varphi_{0}),
\]
where $\varphi_{0}=2\alpha_{\rm R}d_{\rm m}$
and is independent of the dc current bias. 
This expression indicates that the first-harmonic component contains both sine and cosine contributions, 
which cannot be reduced to a simple sine function.
The phase shift originates from the coexistence of a finite exchange field and a finite ${\tilde \alpha}$, as shown in Eq.~(\ref{jjy1}).
The resulting sizable phase shift in the FHSJC plays a crucial role in realizing the SJDE.

\begin{figure}[!h]
 \centering
        \includegraphics[width=0.45\textwidth]{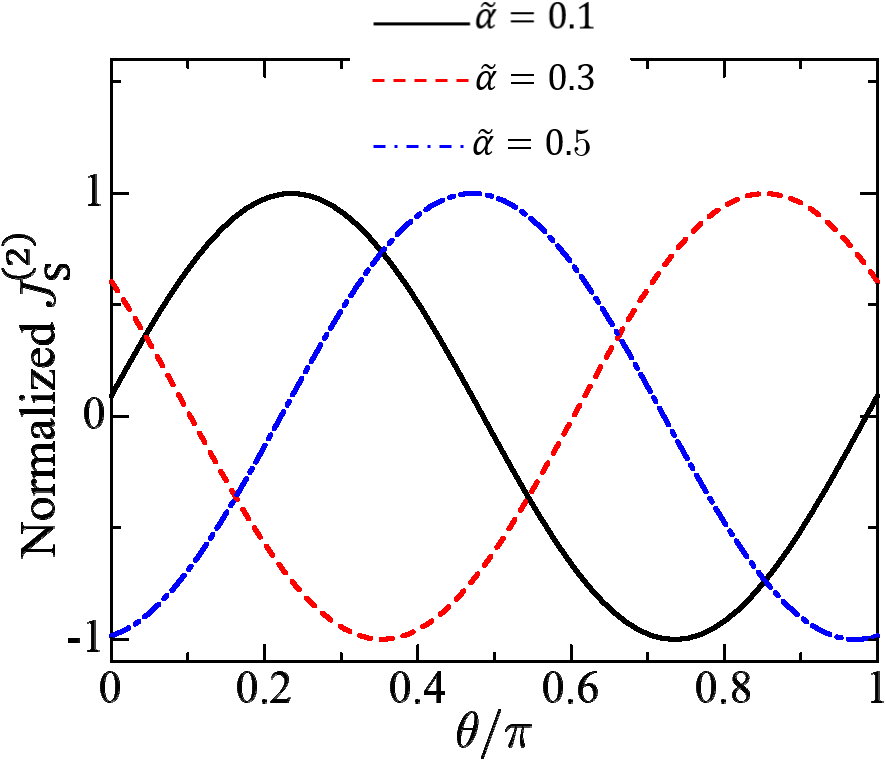}
 \caption{
The spin current--phase relation (SCPR) of the second-harmonic spin Josephson current (SHSJC) for $\tilde{\alpha}=0.1$, $0.3$, and $0.5$.
The SHSJC is normalized by its maximum value.
A sizable phase shift appears in the SCPR depending on $\tilde{\alpha}$.
The coexistence of the second-harmonic contribution and the $\varphi_0$-phase shift constitutes the microscopic origin of the SJDE in the present junction.
}
\label{fig6}
\end{figure}
Figure~\ref{fig6} shows the SCPR of the SHSJC
for ${\tilde \alpha}=0.1$, $0.3$, and $0.5$.
The vertical axis represents the SHSJC normalized such that its maximum value is unity,
while the horizontal axis denotes the superconducting phase difference between the two superconductors. 
It is clearly seen that the SHSJC exhibits a sizable phase shift in the SCPR depending on ${\tilde \alpha}$.
More precisely, the SHSJC is expressed as 
\[
J^{y,(2)}(d_{\rm m})
\propto
\sin (2\theta + 2\varphi_{0})
- 3 \cos (2\theta + 2\varphi_{0}),
\]
which indicates that the second-harmonic component contains both sine and cosine contributions. 
This structure cannot be reduced to a simple phase-shifted sine function. 
The phase shift of the SHSJC is effectively twice that of the FHSJC, namely $2\varphi_{0}$.
Such a mixture of sine and cosine components breaks the antisymmetry of the SCPR and plays a fundamental role in realizing the SJDE. 
In the next subsection, we assess the efficiency of the SJDE within the present approximated framework.

\subsection*{Numerical results of the efficiency of the spin Josephson diode effect}
In this subsection, we numerically assess the efficiency of the SJDE.
To evaluate the efficiency, we consider the total spin Josephson current
defined as
\begin{equation}
J_{\rm s}(d_{\rm m}) = J^{y,(1)}(d_{\rm m}) + J^{y,(2)}(d_{\rm m}),
\label{js}
\end{equation}
where $J^{y,(1)}(d_{\rm m})$ and $J^{y,(2)}(d_{\rm m})$
are given by Eqs.~(\ref{jjy1}) and (\ref{jjy2}), respectively.

We define the efficiency of the SJDE as
\begin{equation}
\eta_{\rm s}
=
\frac{
J_{\rm s}^{\rm max} - \left| J_{\rm s}^{\rm min} \right|
}{
J_{\rm s}^{\rm max} + \left| J_{\rm s}^{\rm min} \right|
},
\label{eta}
\end{equation}
where $J_{\rm s}^{\rm max}$ and $J_{\rm s}^{\rm min}$ denote the maximum and minimum values
of $J_{\rm s}$ with respect to the phase difference $\theta$, respectively.
Here, $J_{\rm s}^{\rm max}$ and $J_{\rm s}^{\rm min}$ correspond to the critical spin currents
in the forward and backward directions, respectively. 
The absolute value is taken for $J_{\rm s}^{\rm min}$ in order to compare
the magnitudes of the spin currents for opposite bias directions on equal footing.
This definition quantifies the degree of nonreciprocity in the SCPR,
which originates from the $\varphi_{0}$ phase shift and higher-harmonic contributions.
In this measure, $\eta_{\rm s}=0$ corresponds to a reciprocal current, while $\eta_{\rm s}=1$ indicates an ideal diode behavior.

\begin{figure}[!t]
 \centering
        \includegraphics[width=0.45\textwidth]{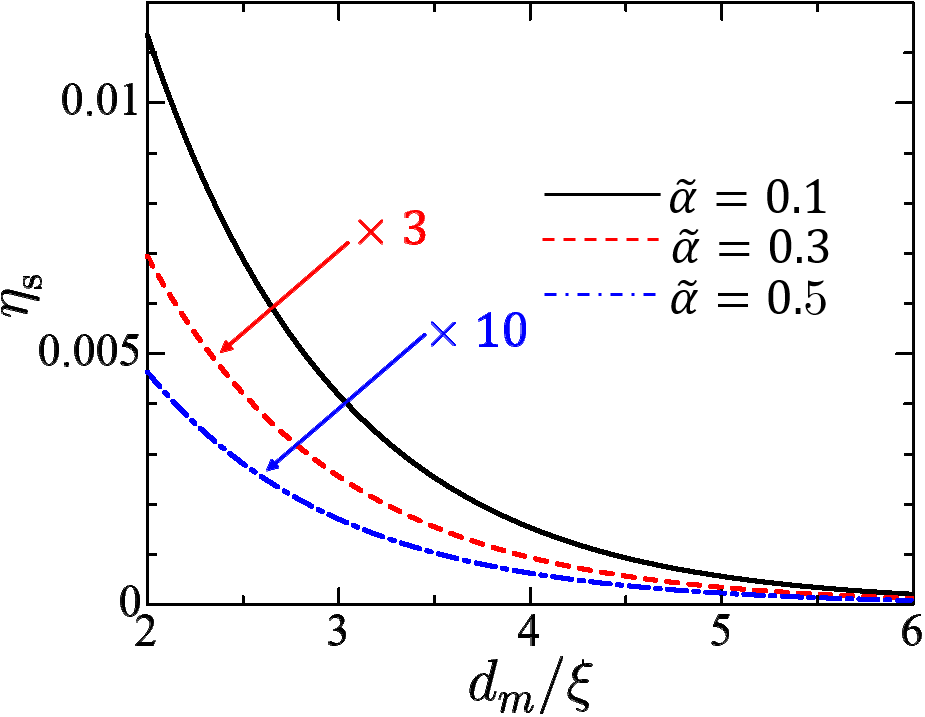}
 \caption{
The efficiency of the spin Josephson diode effect ($\eta_{\rm s}$) as a function of the thickness of the Rashba metal ($d_{\rm m}$) for $\tilde{\alpha}=\xi\alpha_{\rm R}=0.1$, $0.3$, and $0.5$.
$\eta_{\rm s}$ decreases monotonically with increasing $d_{\rm m}$ and is enhanced as $\tilde{\alpha}$ decreases. 
Here, $\xi=\sqrt{\hbar D/(2\pi k_{\rm B} T)}$. 
The finite value of $\eta_{\rm s}$ directly demonstrates the emergence of the SJDE in the present junction.
}
\label{fig7}
\end{figure}

\begin{figure}[!t]
 \centering
        \includegraphics[width=0.45\textwidth]{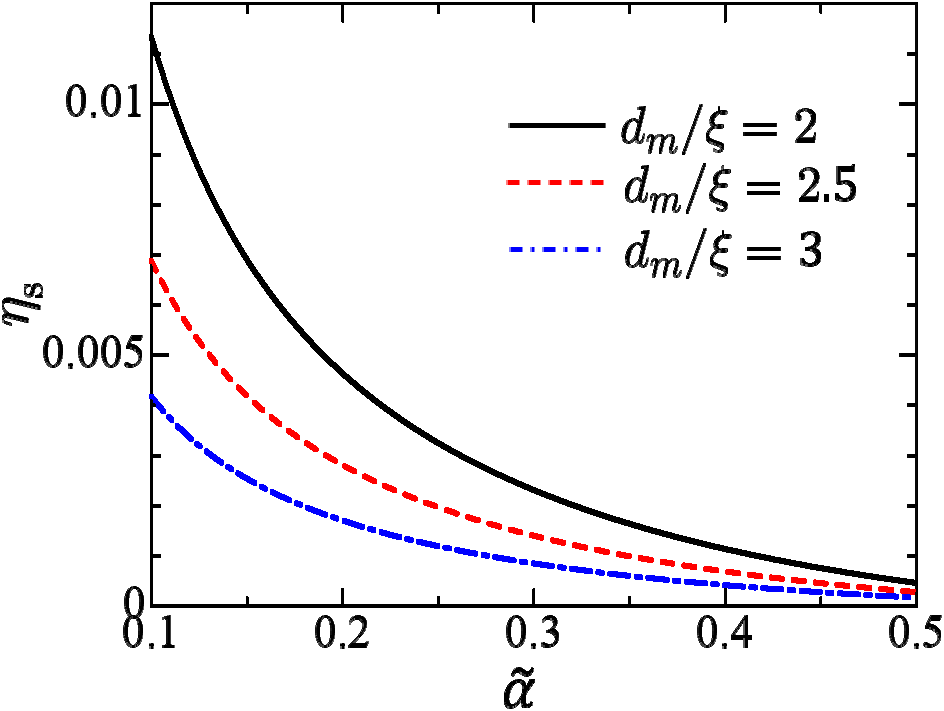}
 \caption{
The efficiency of the spin Josephson diode effect ($\eta_{\rm s}$) as a function of the dimensionless Rashba spin-orbit interaction strength ($\tilde{\alpha}=\xi\alpha_{\rm R}$) 
for $d_{\rm m}/\xi=2$, $2.5$, and $3$.
$\eta_{\rm s}$ increases with decreasing $\tilde{\alpha}$ and decreases as $d_{\rm m}$ increases.
Here, $\xi=\sqrt{\hbar D/(2\pi k_{\rm B} T)}$.
The strong dependence of $\eta_{\rm s}$ on $\tilde{\alpha}$ highlights the essential role of the Rashba spin-orbit interaction in controlling the efficiency of the SJDE.
}
\label{fig8}
\end{figure}

Figures~\ref{fig7} and \ref{fig8} show the numerical results for the efficiency of the SJDE ($\eta_{\rm s}$)
as a function of $d_{\rm m}$ and ${\tilde \alpha}$, respectively.
Figure~\ref{fig7} demonstrates that $\eta_{\rm s}$ decreases monotonically with increasing $d_{\rm m}$ for all values of ${\tilde \alpha}$,
reflecting the rapid suppression of the second-harmonic component, which reduces the asymmetry of the SCPR.
Figure~\ref{fig8} shows that $\eta_{\rm s}$ generally decreases with increasing ${\tilde \alpha}$ for fixed $d_{\rm m}$.
Although increasing ${\tilde \alpha}$ enhances the phase shift in the SCPR,
the magnitude of the second-harmonic component is simultaneously suppressed due to the inverse scaling $J^{y,(2)}(d_{\rm m}) \propto 1/{\tilde \alpha}$.
As a result, the contribution of the second-harmonic component becomes less significant at larger ${\tilde \alpha}$,
leading to a reduction of $\eta_{\rm s}$.

These results indicate that the efficiency of the SJDE is governed by a competition between
the $\varphi_{0}$ phase shift and the second-harmonic amplitude.
In the present parameter range, the suppression of the second-harmonic component dominates,
resulting in a decrease of $\eta_{\rm s}$ with increasing ${\tilde \alpha}$.
Moreover, $\eta_{\rm s}$ is enhanced for smaller $d_{\rm m}$,
where the second-harmonic contribution remains sufficiently large.
Therefore, both a finite second-harmonic component and a moderate phase shift are essential
for realizing a high efficiency SJDE.

\section*{Discussion}
In this section, we discuss the physical implications of the present results, 
with particular emphasis on the origin of the spin Josephson diode effect (SJDE) 
and its distinction from conventional Josephson diode mechanisms, 
and outline possible future research directions. 

The physical origin of the SJDE in the present diffusive junction can be understood as follows. 
Due to the superconducting proximity effect, superconducting pair correlations penetrate into the RM. 
The exchange field induces spin mixing and converts spin-singlet correlations into spin-triplet correlations with zero spin projection. 
It should be emphasized that the RSOI alone does not generate the spin-triplet correlations 
in the present perturbative solution. 
As seen from Eqs.~(26) and (27), 
the exchange field is required to induce the initial spin-triplet component from the spin-singlet pair correlation. 
The RSOI then rotates the spin of the induced triplet correlations 
and gives them additional spin-dependent phases through the effective magnetic field in the RM. 
Subsequently, equal-spin triplet correlations are generated through this spin rotation. 
As a consequence, a finite $\varphi_0$ phase shift appears in the spin current--phase relation. 
Furthermore, higher-order triplet correlations also acquire spin-dependent phases, 
which generate second-harmonic contributions to the spin Josephson current. 
The coexistence of the $\varphi_0$ phase shift and the second-harmonic contribution produces the nonreciprocal spin transport characteristic of the SJDE.

We discuss why $J^{y,(1(2))}(d_{\rm m})$ appears as a purely imaginary quantity 
within the present theoretical framework.
In our formulation, the anomalous Green's function is written as
\begin{eqnarray}
\hat{f}^{(1(2))}(x) &=&
\left(
f^{(1(2))}_{s}(x) + \bm{f}^{(1(2))}(x)\cdot \hat{\bm{\sigma}}
\right)
i \hat{\sigma}_{y}.
\end{eqnarray}
By explicitly expanding the spin structure, we obtain
\begin{eqnarray}
\hat{f}^{(1(2))}(x)
&=&
f_{s}^{(1(2))}(x)
\left(
\begin{array}{cc}
0 & 1 \\
-1 & 0
\end{array}
\right)
+
f_{x}^{(1(2))}(x)
\left(
\begin{array}{cc}
-1 & 0 \\
0 & 1
\end{array}
\right)
+
i f_{y}^{(1(2))}(x)
\left(
\begin{array}{cc}
1 & 0 \\
0 & 1
\end{array}
\right)
+
f_{z}^{(1(2))}(x)
\left(
\begin{array}{cc}
0 & 1 \\
1 & 0
\end{array}
\right).
\label{hat-f-discussion} \nonumber \\
\end{eqnarray}
It is clearly seen from equation~(\ref{hat-f-discussion}) that only $f_{y}(x)$ is accompanied by an explicit imaginary unit,
whereas the other components remain real in this representation.
This feature directly originates from the spin structure of the pairing term $i\hat{\sigma}_{y}$.
Since the spin current is calculated based on Tokatly's SU(2) covariant formulation, 
which involves traces over spin matrices and spatial gradients of 
$\hat{f}$, the imaginary prefactor associated with $f_{y}(x)$ inevitably leads to a purely imaginary expression for $J^{y,(1(2))}(x)$.
In contrast, $J^{x,(1(2))}(x)$ and $J^{z,(1(2))}(x)$ remain real quantities.
Therefore, within the present theoretical framework, 
the purely imaginary nature of $J^{y,(1(2))}(x)$ is an inevitable consequence of the algebraic structure of the anomalous Green's function.
Although $J^{y,(1(2))}(x)$ appears as a purely imaginary quantity within the present representation,
this does not imply any unphysical behavior.
The imaginary factor originates from the specific spin structure of the anomalous Green's function, 
and physical observables are obtained from appropriate real combinations of the Green's functions in the final expressions. 
Therefore, the imaginary nature should be understood as a representation-dependent feature rather than a direct physical observable, 
and does not affect measurable quantities.
In this respect, it would be desirable to formulate the spin current in a manner 
that yields explicitly real-valued expressions at each stage of the calculation. 
Such a formulation, however, is beyond the scope of the present work and remains an interesting issue for future study(also see Appendix A).

In the present junction, the $x$- and $z$-components of the SJC reduce to the first-harmonic dependence,
because the second-harmonic terms in equations~(\ref{jx2}) and (\ref{jz2}) combine with the first-harmonic contributions as discussed in Sec.~3.
As a result, the SJDE does not appear in the $x$- and $z$- components of the SJC within the present configuration. 
In the present configuration, the magnetization is aligned such that no $y$-component of the exchange field is present, 
which suppresses the generation of $f_{y}$. 
As a result, the second-harmonic contributions to the $x$- and $z$-components do not appear, and the SJDE is absent in these components. 
Nevertheless, it is instructive to discuss the possibility of finite $x$- and $z$- contributions to the SJDE. 
For the SJDE to occur, the SJC should contain both first- and second-harmonic terms, which can generally be written as 
\begin{equation}
J^{x(z)}(x)
=
J_{\rm c1}^{x(z)}(x) \sin(\theta+\varphi_0)
+
\tilde{J}_{\rm c1}^{x(z)}(x) \cos(\theta+\varphi_0)
+
J_{\rm c2}^{x(z)}(x) \sin(2\theta+2\varphi_0)
+
\tilde{J}_{\rm c2}^{x(z)}(x) \cos(2\theta+2\varphi_0),
\label{jxz}
\end{equation}
where $J_{\rm c1}^{x(z)}(x)$ and $\tilde{J}_{\rm c1}^{x(z)}(x)$
denote the amplitudes of the sine and cosine components of the first-harmonic contribution, respectively,
while $J_{\rm c2}^{x(z)}(x)$ and $\tilde{J}_{\rm c2}^{x(z)}(x)$ represent those of the second-harmonic contribution. 
These coefficients characterize the relative weights of the even (cosine) and odd (sine) components in the SCPR.
The first-harmonic contributions produce a $\varphi_0$-shifted SCPR, 
while the second-harmonic contributions introduce the $2\theta$ dependence required for the SJDE.
From equations~(\ref{jx2}) and (\ref{jz2}), 
the SHSJC in the $x$- and $z$- components may appear 
when the exchange field acquires a finite $y$ component,
which enables the generation of a finite $f_{y}(x)$.
In this case, $f_{y}(x)$ becomes finite and can contribute to $J^{x,(2)}(x)$ and $J^{z,(2)}(x)$.
Since the magnetization direction can be varied, 
the spin current components responsible for the SJDE may be selectively activated by tuning the magnetization configuration.

This suggests a possibility of controlling the direction of the spin Josephson diode response through magnetization engineering.
Therefore, the directional selectivity of the SJDE is closely related to the magnetization configuration. 
If the magnetization direction is changed so as to generate a sizable $f_{y}(x)$ component, 
the triplet-interference terms appearing in $J^{x,(2)}(x)$ or $J^{z,(2)}(x)$ may also contribute to the SJDE. 
In this sense, the $y$-component is not the only possible active component of the SJDE, 
but rather the component selected by the Rashba spin-orbit coupling and the exchange-field configuration considered in the present work.

We study the first- and second-harmonic SJCs 
and the efficiency of the SJDE ($\eta_{\rm s}$) near the superconducting transition temperature ($T_{\rm c}$).
In general, the amplitude of the SJC is small near $T_{\rm c}$ because the SJC is proportional 
to the superconducting gap, which is strongly suppressed close to $T_{\rm c}$.
As a result, the efficiency $\eta_{\rm s}$ is also reduced in this temperature region.
At lower temperatures, where the superconducting gap becomes larger, both the amplitude of the SJC
and higher-order harmonic contributions are expected to increase.
This suggests that the efficiency $\eta_{\rm s}$ can be significantly enhanced beyond the present high-temperature regime. 
This indicates that the present results provide a conservative estimate of the SJDE efficiency.
Significantly larger efficiencies are expected at lower temperatures.

The efficiency $\eta_{\rm s}$ is calculated under the approximation 
that the thickness of the RM ($d_{\rm m}$) is much larger than the normal-metal coherence length ($\xi$), i.e., $d_{\rm m} \gg \xi$.
From figures~(\ref{fig7}) and (\ref{fig8}),
it is immediately seen that $\eta_{\rm s}$ decreases with increasing $d_{\rm m}$.
This behavior can be attributed to the strong suppression of the SHSJC with increasing $d_{\rm m}$,
compared with the decay rate of the first-harmonic component, as shown in equations~(\ref{jjy1}) and (\ref{jjy2}). 
This reflects the fact that higher-harmonic components are more sensitive to spatial decay and dephasing effects in diffusive systems.
This result suggests that a higher efficiency $\eta_{\rm s}$ can be achieved when the thickness of the RM is sufficiently thin. 

It should be noted that the present theoretical model assumes ideally transparent interfaces between all layers.
In realistic multilayer junctions, however, achieving highly transparent interfaces remains experimentally challenging.
Since the SJDE originates from the interplay between the first- and second-harmonic SJCs,
its magnitude can be sensitive to interface transparency. 
This is because higher-order harmonic processes require coherent multiple scattering,
which is strongly suppressed by interface resistance. 
Moreover, interface roughness introduces additional momentum randomization and dephasing, 
which tends to suppress the second-harmonic spin Josephson current and other higher-order harmonic contributions. 
In particular, since higher-harmonic components are essential for generating nonreciprocity, 
their suppression due to interface scattering may substantially reduce the SJDE efficiency in practical devices. 
Although the SJDE is expected to remain qualitatively present 
as long as both the first- and second-harmonic components survive,
its efficiency is likely to be reduced because 
the higher-harmonic contribution is generally more sensitive to interfacial imperfections than the first-harmonic contribution.

To gain further insight into the origin of the SJDE, we examine the structure of the SJC. 
The conventional Josephson current is typically expressed as a product of anomalous Green's functions with the same spin symmetry (e.g., singlet--singlet correlations). 
In contrast, the SJC intrinsically involves coupling between different spin components.
For instance, the $x$-component of the spin current contains terms such as 
$\bar{f}_{z}\partial_{x} f_{y} - \bar{f}_{y}\partial_{x} f_{z}$ and 
$\bar{f}_{x}\partial_{x} f_{s} - \bar{f}_{s}\partial_{x} f_{x}$, 
which correspond to triplet--triplet and singlet--triplet mixing, respectively. 
This clearly indicates that the SJC arises from the interference between distinct spin sectors, rather than from a single pairing channel.
As a result, the suppression of the spin-singlet component, 
which is often required in conventional JDE systems to generate a sizable $\varphi_{0}$ phase shift~\cite{hikino-jpsj94, hikino-ph}, 
is not a necessary condition for the realization of the SJDE in the present system. 
Instead, the essential ingredient for the SJDE is the coherent coupling 
between different spin components mediated by spin--orbit interaction and the exchange field.
These results demonstrate that the SJDE is not merely a spin analogue of the conventional JDE, 
but originates from a fundamentally different mechanism based on spin-dependent coherence and mixing effects.

The present Josephson junction is constructed from a metallic multilayer with diffusive transport.
In general, it is expected that simpler Josephson junctions with ballistic transport may exhibit a stronger nonlinear
spin current--phase relation and a higher SJDE efficiency. 
In ballistic systems, stronger phase coherence and reduced scattering enhance higher-order harmonic contributions,
leading to a more pronounced nonreciprocal response.
To realize simple Josephson junctions showing the SJDE,
junctions composed of a Rashba ferromagnet in the ballistic transport regime could be a promising candidate,
since the Rashba ferromagnet spontaneously breaks time-reversal and inversion symmetries. 
This suggests that ballistic Rashba ferromagnetic junctions provide a promising platform 
for realizing a robust SJDE without requiring complex magnetic structures.
This issue will be addressed in future work, where a more pronounced SJDE is expected. 

\section*{Summary}
In this work, we have theoretically investigated the first- and second-harmonic spin Josephson currents (FHSJC and SHSJC) and evaluated the spin Josephson diode effect (SJDE) in a diffusive Josephson junction with a Rashba metal layer in the presence of a ferromagnetic exchange field, using the quasiclassical Green's function theory.
Both the FHSJC and the SHSJC exhibit damped oscillatory behavior as a function of the thickness of the Rashba metal ($d_{\rm m}$).
Moreover, the oscillation period becomes shorter with increasing Rashba spin--orbit interaction strength ($\alpha_{\rm R}$).
This oscillatory behavior indicates that the direction of the spin Josephson current can be reversed by tuning $d_{\rm m}$ or $\alpha_{\rm R}$.

The spin current–phase relations (SCPRs) of both the FHSJC and the SHSJC exhibit a finite phase shift 
even in the absence of an external magnetic field. This phase shift originates from additional cosine terms in the FHSJC and the SHSJC. 
The resulting $\varphi_{0}$-shifted SCPRs indicate the breaking of inversion and time-reversal symmetries in the present junction. 
However, the $\varphi_{0}$ phase shift alone is not sufficient to generate the SJDE. 
The second-harmonic spin Josephson current provides an additional phase dependence 
that breaks the symmetry between the forward and backward critical spin currents. 
Therefore, the coexistence of the $\varphi_{0}$ phase shift and the second-harmonic contribution gives rise to the SJDE.

Furthermore, we evaluated the efficiency of the SJDE ($\eta_{\rm s}$) and clarified its dependence on $d_{\rm m}$ and $\alpha_{\rm R}$.
The efficiency decreases with increasing $d_{\rm m}$, reflecting the rapid suppression of the second-harmonic component.
With respect to $\alpha_{\rm R}$, the efficiency is governed by competing effects:
while a larger $\alpha_{\rm R}$ enhances the phase shift in the SCPRs,
it simultaneously suppresses the second-harmonic component due to the inverse scaling of the SHSJC.
As a result, the SJDE efficiency is not determined solely by increasing $\alpha_{\rm R}$, but by a balance between these two effects.

Our results demonstrate that the interplay between the FHSJC and the SHSJC, together with the phase shift in the SCPRs,
provides a mechanism for generating nonreciprocal spin Josephson transport.
This mechanism allows the SJDE to occur even in the absence of an external magnetic field,
and highlights the essential role of higher-order harmonic components in realizing nonreciprocity.

\section{Appendix A. Physical Interpretation of the Spin-Current Components}
The spin-current components contain contributions originating from both singlet-triplet coupling 
and interference between different triplet correlations, as described in Section~3. 
In particular, the latter plays an essential role in the present system and is governed by 
the relative phases between different triplet pairing correlations. 
The peculiar form of the $y$-component does not imply a different physical mechanism. 
Instead, it originates from the purely imaginary character of the Pauli matrix $\sigma_y$, 
whereas $\sigma_x$ and $\sigma_z$ are real matrices. 
Therefore, the formal imaginary structure appearing in $J_y(x)$ should be understood 
as a consequence of the spin algebra rather than a distinct physical origin of the spin current. 
In this sense, all three spin-current components share the same physical origin, 
while their different mathematical forms arise from the algebraic properties of the corresponding Pauli matrices.

\section*{Conflicts of interest}
The author declares no competing interests.

\section*{Funding}
This research received no specific grant from any funding agency.

\section*{Data availability}
No new data were created or analyzed in this study.


\end{document}